\documentstyle[aas2pp4]{article}

\def\simless{\mathbin{\lower 3pt\hbox
     {$\rlap{\raise 5pt\hbox{$\char'074$}}\mathchar"7218$}}}   
\def\simmore{\mathbin{\lower 3pt\hbox
     {$\rlap{\raise 5pt\hbox{$\char'076$}}\mathchar"7218$}}}   
\def\msun{{\rm M}_\odot}                                       

\received{}
\accepted{}
\slugcomment{Submitted to Ap.J.Letters., January 1998}

\lefthead{}
\righthead{}

\begin{document}

\title{``Canonical'' Black Hole States in the Superluminal
       Source GRO~J1655--40}

\author{Mariano M\'endez\altaffilmark{1,2},
        Tomaso Belloni\altaffilmark{1},
        \and
        Michiel van der Klis\altaffilmark{1}
}

\altaffiltext{1}{Astronomical Institute ``Anton Pannekoek'',
       University of Amsterdam and Center for High-Energy Astrophysics,
       Kruislaan 403, NL-1098 SJ Amsterdam, the Netherlands}

\altaffiltext{2}{Facultad de Ciencias Astron\'omicas y Geof\'{\i}sicas, 
       Universidad Nacional de La Plata, Paseo del Bosque S/N, 
       1900 La Plata, Argentina}

\begin{abstract}

We analyze RXTE/PCA observations of the black hole candidate and
galactic superluminal source GRO~J1655--40 during its recent outburst.
We show that during its decay to quiescence, GRO~J1655--40 goes through
the high, intermediate, and low state (and that at the beginning of its
decay it might have even shown signatures of a very high state), just
like other black hole candidates.  This is the first time that such a
transition is observed in a galactic superluminal source.  We discuss
what are the implications of these results on the hypothesis that the
spin of the black hole in superluminal sources is much higher than in
other black hole candidates.

\end{abstract}


\section{Introduction}

Our understanding of low magnetic-field accreting neutron stars and
black hole candidates (BHC) has greatly improved in recent years, and a
general scenario has been proposed in which source properties are
determined by basic parameters, such as accretion rate, compact object
mass, and magnetic field (van der Klis \cite{vanderklis94a}).

One of the keys to this unified scenario has been the successful use of
simultaneous timing and energy spectral data to identify the different
source states, both in neutron star and black hole systems.  For BHCs
four states have been recognized:  (1) the low state (LS; Tananbaum et
al.  \cite{tananbaum72}; Oda et al.  \cite{oda71}), with a flat (photon
index $\Gamma = 1.5-2$) power law X-ray spectrum and strong ($25-50$\,\%
rms) band limited noise with a break frequency of $0.03-0.3$~Hz; (2) an
intermediate state (IS; M\'endez \& van der Klis \cite{mendez97};
Belloni et al.  \cite{betal97}) where the $2-10$~keV flux is a factor of
$\sim 4$ higher than in the LS, a soft component is present in the
energy spectrum while the power law component is softer than in the LS
($\Gamma = 2-3$), the power spectrum shows $6-8$~Hz Quasi-Periodic
Oscillations (QPO), and the break frequency of the band limited noise is
in the range $1-10$~Hz; (3) a high state (HS; Tananbaum et al.
\cite{tananbaum72}; Oda et al.  \cite{oda77}) where the $2 - 10$~keV
flux is an order of magnitude higher than in the LS due to the presence
of an ultrasoft X-ray spectral component with sometimes a $\Gamma = 2-3$
power law tail, and a weak (few \% rms) flat power law power spectrum;
and (4) a very high state (VHS; Miyamoto et al.  \cite{miyamoto91})
characterized by high $2-20$~keV X-ray luminosity ($2-8$ times higher
than in the HS), an ultrasoft X-ray spectral component plus a ($\Gamma
\sim 2.5$) power law tail, strongly variable $1-15$\,\% rms band limited
noise with a much higher cut off frequency ($1-10$~Hz) than in the LS,
and often $3-10$~Hz QPO.  Apart from the difference in $2-20$~keV
luminosity, the timing and spectral properties of the IS are quite
similar to those of the VHS; the distinction comes from the fact that
they are separated by a HS with very different properties.  So far, the
VHS has only been observed in GX~339--4 (Miyamoto et al.
\cite{miyamoto91}) and GS~1124--68 (Ebisawa et al.  \cite{ebisawa94}).

The order in which black hole transients in their decay, and the
persistent BHCs, have been observed to move through these states, and
similarities to neutron star states strongly suggest that the accretion
rate is highest in the VHS, and decreases through the HS, IS, and LS
(Miyamoto et al.  \cite{miyamoto91}; van der Klis \cite{vanderklis94a};
M\'endez \& van der Klis \cite{mendez97}; Belloni et al.
\cite{betal97}).

Two remarkable BHCs, GRS~1915+105 and GRO J1655--40, so far stood out
from the other ones.  They show relativistic ejections, resembling a
small scale analogue to the jets observed in active galactic nuclei and
quasars (Mirabel \& Rodr\'{\i}guez \cite{mr94}), and several other types
of unique behavior (Morgan, Remillard, \& Greiner \cite{mrg97}; Belloni
et al.  \cite{bmk97b}; Fender et al.  \cite{fender97}), but until now
there was little evidence for any of the canonical black hole states
described above.

GRO~J1655--40 was discovered during an outburst in 1994, and since then
it has shown irregular outburst activity (e.g., Zhang et al.
\cite{zes97}).  The most recent X-ray event of this source started on
1996 April 25 (Remillard et al.  \cite{remiau96}), and lasted for about
one and a half year.  Dynamical measurements suggest the compact star in
GRO~J1655--40 is a black hole, with a mass of $\sim 7$~M$_{\odot}$
(Orosz \& Bailyn \cite{ob97}, van der Hooft et al.  \cite{hha98}).  A
few weeks after the 1994 X-ray outburst, radio jets were detected
expanding at apparent superluminal velocities (Tingay et al.
\cite{tingay95}; Hjellming \& Rupen \cite{hr95}).

Here we present an analysis of the properties of GRO~J1655--40 during
the decay from its latest X-ray outburst.  We find, for the first time
in a superluminal source, that GRO~J1655--40 goes through the HS, IS and
LS just like other BHCs.

\section{Observations and Data Analysis}

The Proportional Counting Array (PCA) on board the Rossi X-ray Timing
Explorer (RXTE) observed GRO~J1655--40 regularly from March until
September 1997, when it faded into quiescence.  We selected 8
observations (see Table~\ref{tabspe}), from a list of about 30 publicly
available in this period, that represent a good sample of its decay from
$\simmore 16500$~PCA~c/s ($2-60$~keV) to quiescence (see
Figure~\ref{figasm}).

In all cases data were collected in a high time (and low energy)
resolution mode, simultaneously with a mode with 16-s time resolution in
129 energy channels which covered the full PCA energy band.  We divided
the high time resolution data in segments of 256~s, and produced an
average power density spectrum (PDS) per observation extending from
$1/256$ to $1024$~Hz.  We subtracted the Poisson noise and the Very
Large Event window contribution (Zhang et al.  \cite{zhang95i}; Zhang
\cite{zhang95ii}) from each PDS, and renormalized them to fractional rms
squared per Hertz (see van der Klis \cite{vanderklis95}).  The results
of the fits to these PDS are shown in Table~\ref{tabspe} and
Figure~\ref{figpow}.

On July 29 the PDS fitted a power law with an index of $\sim 1$.  Before
that, on May 28 and July 8, the PDS were very similar to that of July
29, however a single power law did not fit the data well, specially
between 0.1 and 1 Hz.  We fitted these PDS using a broken power law,
although a power law with an exponential cut-off, or two Lorentzians and
a power law (as shown in Figure~\ref{figpow}) fitted as well.  On August
3 the source showed no evidence for variability in our power spectra,
with an upper limit on the fractional rms amplitude consistent with that
of July 29.

Between August 3 and 14 the PDS changed dramatically.  On August 14 and
18 the PDS showed a strong band limited noise component (BLN), a break
at $\simless 1$~Hz whose frequency decreased as the count rate
decreased, and a QPO feature moving from $\sim 6.5$~Hz to $\sim 0.8$~Hz
in correlation with the break of the BLN component.  On August 25 and
29, at much lower count rate, we could not measure any variability from
the source, with upper limits that were consistent with the measurements
of August 14 and 18.  During the next PCA pointing, on September 1, we
could not detect the source above the background level.

On August 14, while no other power spectral component showed a
significant energy dependence, the amplitude of the 6~Hz QPO was lower
at low energies, the rms amplitudes below and above 5~keV being $8.4 \pm
0.3$\,\%, and $11.3 \pm 0.2$\,\%, respectively.  On August 18 none of
the fitted parameters depended significantly on energy.

We used the 16-s data to produce $2-30$~keV energy spectra for each
observation which we fitted with a model consisting of a disk-blackbody
(DBB, Mitsuda et al.  \cite{mal84}) component plus a power law.  We also
added a Gaussian line plus an edge to fit the data around 7~keV, but the
other parameters were not sensitive to the inclusion of these
components.  The interstellar absorption was kept fixed at $8.9 \times
10^{21}$ cm$^{-2}$ (Zhang et al.  \cite{zes97}).  The best fit
parameters are shown in Table~\ref{tabspe}.

During these observations the temperature of the DBB decreased
monotonically from $\sim 1.2$~keV to $\sim 0.5$~keV but the derived
inner radius of the disk did not change significantly.  While the
$2-10$~keV flux decreased steadily during these observations, the
relative contribution of the DBB component remained more or less
constant at $\sim 90$\,\% until August 3.  Between August 3 and 14 this
contribution dropped considerably; from August 14 onwards it is
$\simless 30$\,\%.

\section{Discussion}

Our results show that the BHC and superluminal source GRO~J1655--40 fits
in with the general picture of BHC states described in Section 1.
During these observations the total $2-10$~keV flux decayed by a factor
of $\simmore 5000$, the power law component in the energy spectrum
became harder, and the fractional contribution of the soft component to
the total flux decreased by a factor of $\sim 3$.  Simultaneously, the
time variability increased by a factor of $\simmore 3$, and the power
spectrum developed a flat-topped band-limited noise component with a
break frequency of $\sim 0.1 - 1$~Hz.  All these changes, both in the
energy and power spectra, indicate that GRO~J1655--40 gradually passed
through the HS, IS and LS.  This is the first time that such a
transition was observed in a galactic superluminal source, and it
provides a unique opportunity to understand these sources in terms of
canonical black hole states.

One of the characteristic properties of the band limited noise component
of BHC in the LS and IS (and the VHS) is its variable break frequency,
which anticorrelates with the rms amplitude measured at the break
(Belloni \& Hasinger \cite{b&h90}, M\'endez \& van der Klis
\cite{mendez97}).  The PDS of August 14 and 18 show the same
anticorrelation and fit in with previously observed sources (see
M\'endez \& van der Klis \cite{mendez97}), supporting the conclusion
that GRO~J1655--40 behaves similarly to other BHCs.

On July 29 and August 3 the PDS are consistent with those of a typical
HS (van der Klis \cite{vanderklis94a}), but on May 28 and July 8 they do
not fit the canonical $\alpha \sim 1 - 1.5$ single power law shape.  Due
to the small rms amplitude that characterizes it, the PDS in the HS was
not very well constrained in most previous observations.  It may well be
that the ``unusual'' shape of the PDS on May 28 and July 8 (compared to
what is expected in the HS) is the rule, but it is only the greater
sensitivity of RXTE that allows us to measure it in more detail.

On May 28 the rms amplitude is also larger than was typically observed
in the HS, and comparable to that of the ``anomalous'' state reported by
Kuulkers, van der Klis, \& Parmar (\cite{kvp97}) in 4U~1630--47.
Perhaps this is an aspect of VHS behavior.  The May 28 $2-10$ flux is
$\simmore 60$ times higher than in the LS (August 18), and 4 times
higher than in the HS (July 29), and both the shape of the power
spectrum and the total rms variability are very similar to those
observed in GS~1124--68 on February 6 1991, during its VHS (compare the
power spectrum in Figure~\ref{figpow} with that of Figure~1 in Miyamoto
et al.  \cite{miyamoto94}).  This would make GRO~J1655--40 the third BHC
(after GX~339--4 and GS~1124--68) to be observed in the VHS.

The application of the DBB plus power law model to BHCs, previously
strongly advocated on the basis of Ginga observations of classical BHC
(Tanaka \& Lewin \cite{tl95}) gains credence by the link provided by our
observations between the superluminal and the classical BHC.  In
GRS~1915+105, the application of this model leads to a very compelling
explanation (Belloni et al.  \cite{bmk97b}) for the large luminosity
excursions in terms of rapid variations of the inner disk radius.

\begin{figure}[h]
\plotfiddle{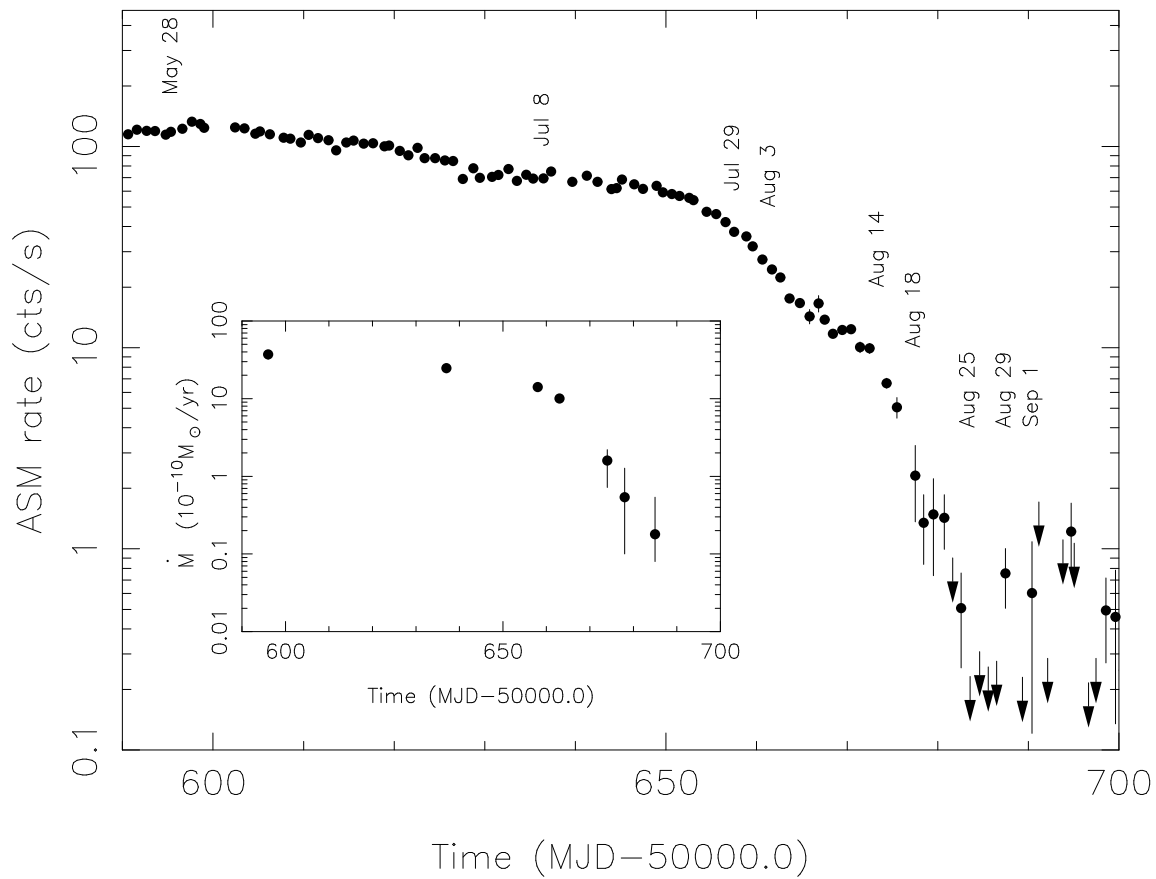}{250pt}{270}{100}{100}{-300}{400}
\vspace{5cm}
\caption{RXTE All-Sky Monitor light curve (one-day
averages) of GRO~J1655--40 from 1997 May 22 to 1997 September 9.
Background rates were estimated based on the data points after this
date, when the source is not detected by the ASM, and subtracted.
Inset: Inferred mass accretion rate for the observations of May 28,
July 8, 29, August 3, 14, 18, and 25.
\label{figasm}}
\end{figure}

\begin{figure}[h]
\plotfiddle{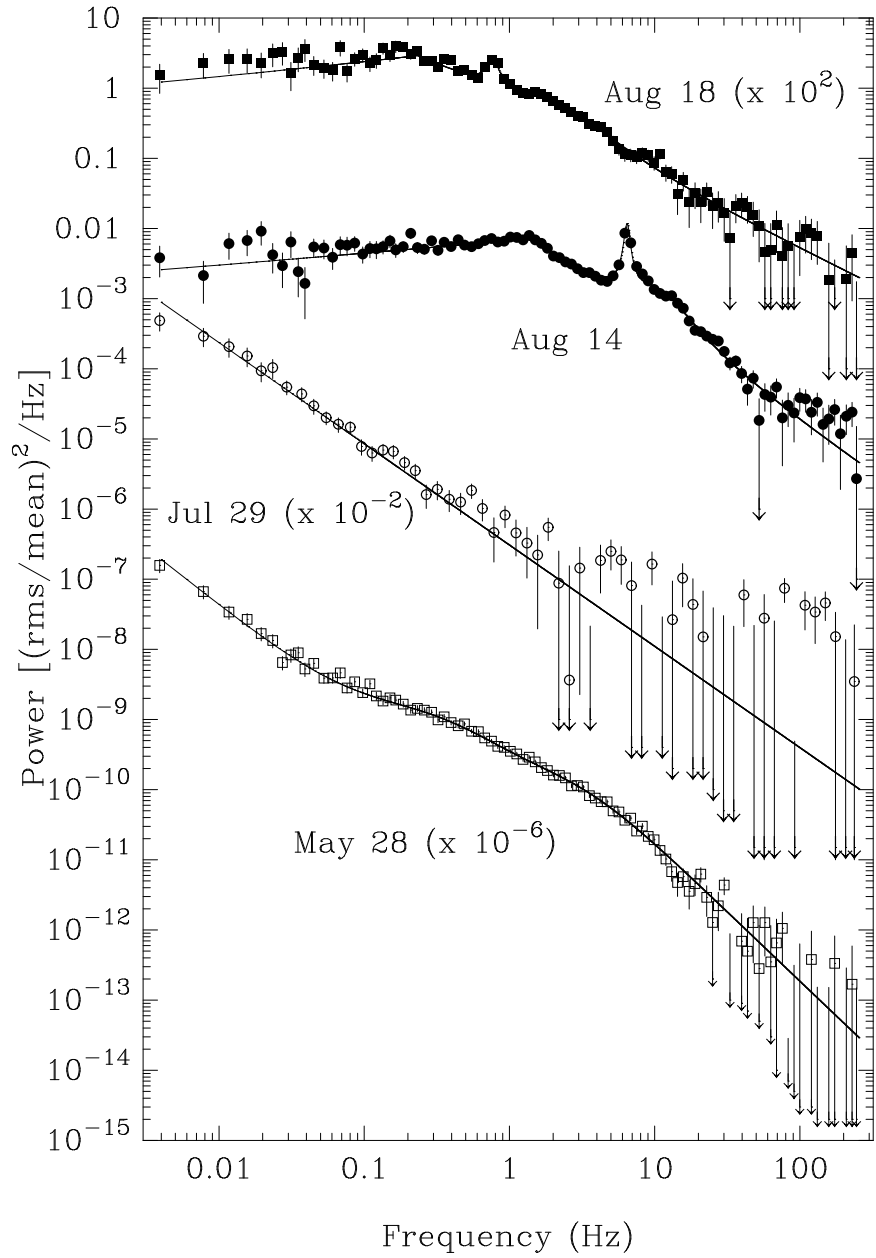}{250pt}{0}{70}{70}{-200}{-200}
\vspace{8cm}
\caption{The power spectra of the observations of May 28,
July 29, August 14, and 18.  The total fractional rms amplitudes are
4.7\,\%, 1.8\,\%, 15.6\,\%, and 18.0\,\%, respectively.
\label{figpow}}
\end{figure}

Using the values of $R_{\rm in}$ and $T_{\rm in}$ obtained from the
spectral fits, and the derived mass of the BHC (Orosz \& Bailyn
\cite{ob97}; van der Hooft et al.  \cite{hha98}), we can infer the
accretion rate of GRO~J1655--40 during each observation (see
Figure~\ref{figasm}, inset).

Despite the large decrease of $\dot M$, the DBB flux and the total flux
during this transition, the inner radius of the disk remained the same.
The fitted values are consistent with a constant radius of 24.8~km.  The
same has been observed in GS~1124--68, GS~2000+25, GX~339--4 and
LMC~X--3 (Tanaka \& Lewin \cite{tl95}, and references therein), although
in those cases the flux decreased by less than two orders of magnitude,
not so dramatically as in GRO~J1655--40.  Tanaka (\cite{tan92})
interpreted this lack of change in $R_{\rm in}$ as a signature of the
innermost stable orbit around a black hole.  If we assume that the value
of $R_{\rm in} = 24.8$~km that we obtain from the fits is in fact the
radius of the innermost stable orbit, the required spin parameter for a
7~$\msun$ black hole is $a_{\rm *} = +0.89$.

Recently, Cui, Zhang, \& Cheng (\cite{czc98}) proposed that certain QPO
features observed in several BHC are produced by the precession of the
accretion disk due to the relativistic dragging of the inertial frame
around a spinning black hole.  Using the 300~Hz QPO in GRO~J1655--40
(Remillard et al.  \cite{remetal97}), and the 67~Hz QPO in GRS~1915+105
(Morgan et al.  \cite{mrg97}), they conclude that both galactic
superluminal sources contain a very rapidly rotating black hole ($a_{\rm
*} \sim +0.95$).  Similarly, they interpret the 6.7~Hz peak in
GS~1124--68 (Belloni et al.  \cite{betal97}) and the 8~Hz QPO in
Cyg~X--1 (Cui et al.  \cite{cuietal97b}) as the same phenomenon, and
conclude that the black holes in these two sources are spinning less
rapidly ($a_{\rm *} \sim +0.35$, $a_{\rm *} \sim +0.48$, respectively).
Based on these results, Cui et al.  (\cite{czc98}) propose that the
difference between BHC that are also superluminal jet sources and
otherwise ``normal'' BHC is the spin of the black hole.

Both in GS~1124--68 and Cyg~X--1, the QPOs used by Cui et al.  to reach
this conclusion were observed during a source transition through the
intermediate state, identical to the one in GRO~J1655--40 that we
describe here.  Remarkably, on August 14 and 18 the PDS show a QPO
feature moving from $\sim 6.5$~Hz to $\sim 0.8$~Hz, which is very
similar to those observed in GS~1124--68 and Cyg~X--1.  In all three
sources the QPO is more prominent at higher energies, and both in
GRO~J1655--40 and in Cyg~X--1 the frequency of the QPO is correlated to
the break frequency of the band limited noise component.

However, if we, as Cui et al.  (\cite{czc98}) did in the case of
GS~1124--68 and Cyg~X--1, interpret this intermediate-state QPO in
GRO~J1655--40 as the precession frequency of the disk, we would derive a
spin of the black hole of +0.35 on August 14, and $\sim -0.1 - + 0.1$ on
August 18 (see Cui et al.  \cite{czc98}, Figure~2), in strong
contradiction with the value of +0.89 from the spectral fits.  This
argues against the interpretation of the intermediate-state QPO as the
precession frequency of the disk.  Of course, these results cannot be
used to rule out the possibility that the 300~Hz and 67~Hz features in
GRO~J1655--40 and GRS~1915+105 are related to the frame dragging effect,
but if they are, then our results provide a strong objection to a
similar interpretation of the 6.7~Hz and 8~Hz QPOs in GS~1124--68 and
Cyg~X--1.  Either way, we must conclude that the hypothesis that the
angular momentum of the black holes in superluminal sources is much
higher than that of other BHC is no longer supported by the observed
differences in QPO properties between GRO~J1655--40 on the one hand, and
Cyg~X--1 and GS~1124--68 on the other.

Finally, we can compare our results for GRO~J1655--40 to those obtained
for the other galactic superluminal source, GRS~1915+105.  By means of
time-resolved spectral analysis, Belloni et al.  (\cite{bmk97b}) were
able to explain all spectral changes observed in GRS~1915+105 as the
effect of thermal-viscous instabilities in the radiation-pressure
dominated region of an accretion disk.  Based on the fit results to
different parts of the highly variable X-ray light curve, they infer
values of $\dot M$ of $\sim 1.7-10 \times 10^{-8} \msun$~yr$^{-1}$
during the quiescent phase (the range being determined by the
uncertainty in the spin of the central black hole), and a factor of two
higher during the so-called burst phase.  The accretion rates estimated
here for GRO~J1655--40 are about an order of magnitude lower than those
of GRS~1915+105 during its quiescent phase.  It is possible that the
complicated variability in the X-ray light curve of GRS~1915+105, which
is not observed in GRO~J1655--40, is entirely due to the higher
accretion rate in that source (although a higher mass, and therefore
Eddington luminosity of GRS~1915+105, may compensate this).  A natural
consequence of this hypothesis is that, at higher accretion rates
GRO~J1655--40 (and other BHCs) should display a similar behavior as the
one observed in GRS~1915+105, while on the other hand, at lower
accretion rates GRS~1915+105 should undergo a source transition, similar
to the one described here for GRO~J1655--40.

\acknowledgements

This work was supported in part by the Netherlands Foundation for
research in astronomy (ASTRON) under grant 781-76-017.  MM is a fellow
of the Consejo Nacional de Investigaciones Cient\'{\i}ficas y T\'ecnicas
de la Rep\'ublica Argentina.

\clearpage

\begin{deluxetable}{lcccccccc}
\tablecolumns{9}
\scriptsize
\tablecaption{Power and X-ray Spectral Parameters
\label{tabspe}}
\tablewidth{0pt}
\tablehead{
\colhead{}                                    &
\colhead{May 28}                              &
\colhead{Jul  8}                              &
\colhead{Jul 29}                              &
\colhead{Aug  3}                              &
\colhead{Aug 14}                              &
\colhead{Aug 18}                              &
\colhead{Aug 25}                              &
\colhead{Aug 29}                              \\
}
\startdata
UTC Start                                     &
07:41                                         &
11:28                                         &
08:26                                         &
15:57                                         &
10:33                                         &
13:37                                         &
10:26                                         &
03:38                                         \\
UTC End                                       &
10:11                                         &
15:35                                         &
09:27                                         &
16:08                                         &
11:22                                         &
14:35                                         &
11:25                                         &
04:03                                         \\
Rate\tablenotemark{a}\ (c/s)                  &
16900                                         &
9100                                          &
4000                                          &
1800                                          &
1100                                          &
320                                           &
40                                            &
4                                             \\
\cutinhead{Power Spectra}
rms$_{\rm BLN}$\tablenotemark{b}\ (\%)        &
$\phm{0}4.7 \pm 0.1\phm{0}$                   &
$\phm{0}2.8 \pm 0.1\phm{0}$                   &
$\phm{0}1.8 \pm 0.1\phm{0}$                   &
$< 2$                                         &
$\phm{-}15.6 \pm 0.4\phm{0}$                  &
$\phm{-}18.0 \pm 0.8\phm{0}$                  &
$< 19$                                        &
$< 84$                                        \\
$\alpha_{1}$                                  &
$0.93 \pm 0.01$                               &
$0.94 \pm 0.01$                               &
$1.00 \pm 0.05$                               &
1\tablenotemark{c}\                           &
$-0.19 \pm 0.03$                              &
$-0.25 \pm 0.10$                              &
1\tablenotemark{c}\                           &
1\tablenotemark{c}\                           \\
$\alpha_{2}$                                  &
$1.70 \pm 0.04$                               &
$1.47 \pm 0.06$                               &
\nodata                                       &
\nodata                                       &
$\phm{-}1.41 \pm 0.05$                        &
$\phm{-}1.02 \pm 0.03$                        &
\nodata                                       &
\nodata                                       \\
$\nu_{\rm break}$ (Hz)                        &
$3.12 \pm 0.17$                               &
$1.21 \pm 0.17$                               &
\nodata                                       &
\nodata                                       &
$\phm{-}1.34 \pm 0.03$                        &
$\phm{-}0.23 \pm 0.02$                        &
\nodata                                       &
\nodata                                       \\
rms$_{\rm QPO_1}$ (\%)                        &
\nodata                                       &
\nodata                                       &
\nodata                                       &
\nodata                                       &
$\phm{-0}9.8 \pm 0.2\phm{0}$                  &
$\phm{-0}5.6 \pm 0.5\phm{0}$                  &
\nodata                                       &
\nodata                                       \\
FWHM (Hz)                                     &
\nodata                                       &
\nodata                                       &
\nodata                                       &
\nodata                                       &
$\phm{-}0.63 \pm 0.03$                        &
$\phm{-}0.16 \pm 0.03$                        &
\nodata                                       &
\nodata                                       \\
$\nu$ (Hz)                                    &
\nodata                                       &
\nodata                                       &
\nodata                                       &
\nodata                                       &
$\phm{-}6.46 \pm 0.01$                        &
$\phm{-}0.77 \pm 0.01$                        &
\nodata                                       &
\nodata                                       \\
rms$_{\rm QPO_2}$ (\%)                        &
\nodata                                       &
\nodata                                       &
\nodata                                       &
\nodata                                       &
$\phm{-}13.2 \pm 0.7\phm{0}$                  &
$\phm{-}15.4 \pm 2.6\phm{0}$                  &
\nodata                                       &
\nodata                                       \\
FWHM (Hz)                                     &
\nodata                                       &
\nodata                                       &
\nodata                                       &
\nodata                                       &
$\phm{-}10.9 \pm 0.8\phm{0}$                  &
$\phm{-0}3.7 \pm 0.5\phm{0}$                  &
\nodata                                       &
\nodata                                       \\
$\nu$ (Hz)                                    &
\nodata                                       &
\nodata                                       &
\nodata                                       &
\nodata                                       &
$\phm{-0}9.3 \pm 0.3\phm{0}$                  &
$\phm{-0}1.3 \pm 0.4\phm{0}$                  &
\nodata                                       &
\nodata                                       \\
$\chi^2$/dof                                  &
380/336                                       &
334/336                                       &
342/338                                       &
\phm{0}89/95\phm{0}                           &
\phm{0}397/330                                &
\phm{0}288/330                                &
379/367                                       &
274/277                                       \\
\cutinhead{Energy Spectra}
$kT_{\rm in}$ (keV)                           &
$1.20 ^{+0.01}_{-0.01}$                       &
$1.05 ^{+0.01}_{-0.01}$                       &
$0.86 ^{+0.01}_{-0.01}$                       &
$0.79 ^{+0.01}_{-0.01}$                       &
$0.46 ^{+0.05}_{-0.04}$                       &
$0.37 ^{+0.04}_{-0.04}$                       &
$0.38 ^{+0.02}_{-0.05}$                       &
\nodata                                       \\
$R_{\rm in}$\tablenotemark{d} (km)            &
$\phm{.00}23 ^{+1\phm{000.}}_{-1\phm{000.}}$  &
$\phm{.00}24 ^{+1\phm{000.}}_{-1\phm{000.}}$  &
$\phm{.00}26 ^{+1\phm{000.}}_{-1\phm{000.}}$  &
$\phm{.00}26 ^{+1\phm{000.}}_{-1\phm{000.}}$  &
$\phm{.00}29 ^{+3\phm{000.}}_{-5\phm{000.}}$  &
$\phm{.00}27 ^{+2\phm{000.}}_{-7\phm{000.}}$  &
$\phm{.00}18 ^{+12\phm{00.}}_{-6\phm{000.}}$  &
\nodata                                       \\
$\Gamma$ (photon index)                       &
$2.49 ^{+0.06}_{-0.06}$                       &
$2.01 ^{+0.01}_{-0.01}$                       &
$2.38 ^{+0.12}_{-0.13}$                       &
$2.18 ^{+0.19}_{-0.13}$                       &
$2.07 ^{+0.01}_{-0.01}$                       &
$1.77 ^{+0.02}_{-0.02}$                       &
$2.12 ^{+0.05}_{-0.05}$                       &
$3.04 ^{+0.51}_{-0.44}$                       \\
Flux\tablenotemark{e}                         &
$42.70$                                       &
$24.64$                                       &
$11.75$                                       &
$\phm{0}7.24$                                 &
$\phm{0}2.28$                                 &
$\phm{0}0.58$                                 &
$\phm{0}0.09$                                 &
$<0.009\phm{0}$                               \\
DBB Flux\tablenotemark{e}                     &
$39.66$                                       &
$23.30$                                       &
$10.26$                                       &
\phm{0}6.66                                   &
\phm{0}0.30                                   &
\phm{0}0.05                                   &
\phm{0}0.03                                   &
\nodata                                       \\
$\chi^2$/dof                                  &
\phm{0}28/46\phm{0}                           &
\phm{0}25/46\phm{0}                           &
\phm{0}20/46\phm{0}                           &
\phm{0}34/46\phm{0}                           &
\phm{0}66/48\phm{0}                           &
\phm{0}46/48\phm{0}                           &
\phm{0}57/48\phm{0}                           &
171/52\phm{0}                                 \\
\enddata
\tablenotetext{a}{$2-60$~keV background subtracted count rate for 5 PCA
detectors.}
\tablenotetext{b}{$0.01-100$~Hz fractional rms amplitude.}
\tablenotetext{c}{The slope was kept fixed to estimate the upper limit.}
\tablenotetext{d}{For a distance of 3.2~kpc (Hjellming \& Rupen \cite{hr95})
and an inclination of 70$^{\circ}$ (Orosz \& Bailyn \cite{ob97}; Van der Hooft
et al. \cite{hha98}).}
\tablenotetext{e}{Unabsorbed flux in the 2--10 keV range, in units
of $10^{-9}$~erg~cm$^{-2}$~s$^{-1}$.}
\tablenotetext{}{N$_{\rm H}$ was kept fixed at $8.9 \times 10^{21}$
cm$^{-2}$ (Zhang et al. \cite{zes97}). Quoted errors represent $1 \sigma$
single parameter confidence intervals. Upper limits 95\,\%.}
\end{deluxetable}

\onecolumn

\clearpage

\clearpage

\end{document}